\date{}
\documentclass[aps,pra,showpacs,twocolumn]{revtex4-1}
\usepackage{graphicx,psfrag,amsmath,amssymb,amsfonts,latexsym,color,dcolumn,epsfig}
\begin{document}
\title{Quantum dissipative effects in graphene-like mirrors}
\author{C\'esar D. Fosco$^{1,2}$}
\author{Fernando C. Lombardo$^3$}
\author{Francisco D. Mazzitelli$^{1}$}
\author{Mar\'\i a L. Remaggi$^{1,2}$}
\affiliation{$^1$ Centro At\'omico Bariloche,
Comisi\'on Nacional de Energ\'\i a At\'omica,
R8402AGP Bariloche, Argentina}
\affiliation{$^2$ Instituto Balseiro,
Universidad Nacional de Cuyo,
R8402AGP Bariloche, Argentina}
\affiliation{$^3$ Departamento de F\'\i sica {\it Juan Jos\'e
 Giambiagi}, FCEyN UBA, Facultad de Ciencias Exactas y Naturales,
 Ciudad Universitaria, Pabell\' on I, 1428 Buenos Aires, Argentina - IFIBA}
\date{today}
\begin{abstract} 
We study quantum dissipative effects due to the accelerated motion of a
single, imperfect, zero-width mirror.  It is assumed that the microscopic degrees of freedom
on the mirror are confined to it, like in plasma or graphene sheets.
Therefore, the mirror is described by a vacuum
polarization tensor $\Pi_{\alpha\beta}$ concentrated on a time-dependent
surface.  Under certain assumptions about the microscopic model for the
mirror, we obtain a rather general expression for the Euclidean effective
action, a functional of the time-dependent mirror's  position, in
terms of two invariants that characterize the tensor $\Pi_{\alpha\beta}$.
The final result can be written in terms of the TE and TM reflection
coefficients of the mirror, with qualitatively different contributions
coming from them. We apply that general expression to derive the
imaginary part of the `in-out' effective action, which measures dissipative
effects induced by the mirror's motion, in different models, in particular 
for an accelerated graphene sheet.
\end{abstract}
\maketitle
\section{Introduction}\label{sec:intro}

One of the most remarkable manifestations of the quantum nature of the
electromagnetic (EM) field is the so called `motion induced radiation' or
`Dynamical Casimir Effect' (DCE), whereby  the accelerated motion of a
mirror can make the EM field vacuum to evolve to an excited state, namely,
one containing a non-vanishing number of photons~\cite{moore}. 

Predictions about potentially observable DCE effects have been obtained for
a variety of geometries and systems, and by means of quite different
theoretical tools~\cite{reviews}.  In this paper we concentrate on the
calculation of DCE effects for the case of `imperfect mirrors', by which we
mean those that do not necessarily impose perfect conductor boundary
conditions. 

To conduct this study,  we shall follow our previous work for scalar and
spinorial vacuum fields~\cite{prd07,prd11} in which we used  the
particularly convenient functional approach proposed in Ref.\cite{GK}. 
This approach is based on the introduction  of auxiliary
fields inside the functional integral for the vacuum field, whose role is
to impose the proper boundary conditions for the scalar field on each
mirror.  We shall here use an adapted version of the method, designed to
deal with the case of imperfect mirrors, in the presence of the quantum EM
field. 

In a recent work~\cite{maghrebi}, the DCE for imperfect mirrors has been
analyzed using a scattering approach, for the case of a quantum scalar
field. Here, instead, we will consider the EM case,  and for mirrors that
can be described by means of their vacuum polarization tensors (VPT), which
in turn are assumed to come from the integration of charged microscopic
degrees of freedom constrained to them.  Our results
are therefore applicable, for instance,  to plasma and  graphene sheets.

Formally, we will compute the
Euclidean effective action, and use analytic continuation to obtain the
imaginary part of the real-time `in-out' effective action. The latter is
proportional to the probability of vacuum decay, an effect due to the
mirror's acceleration. 

The paper is organized as follows: in Section~\ref{sec:effect}, we
describe the kind of system that we shall consider, and define the corresponding
effective action, within the framework of a perturbative expansion in
powers of the departure of the mirror from its equilibrium position (a
planar, static configuration).  
We obtain a general expression for the effective action at the second order in
that expansion, in terms of two scalar functions which entirely define the
response functions of the mirror. Moreover, up to this order,  the result can be 
written as an integral that involves the transverse electric (TE) and transverse magnetic (TM)
reflection coefficients of the mirror. 

In Section~\ref{sec:examples} we evaluate the Euclidean effective action
for different examples, discarding terms that do not contribute to the
imaginary part of the vacuum energy (i.e., to the vacuum decay probability)
when rotated back to Minkowski spacetime. The examples considered are
distinguished by the different choices for the mirror's VPT. We consider, in particular,
the VPT corresponding to a graphene sheet described by massless fermions.

We present our conclusions in  Section~\ref{sec:conclusions}.
\section{The model and its Euclidean effective action}\label{sec:effect}
\subsection{The model}
We shall begin by defining here the characteristics of the system and its
geometry, as well as the conventions and approximations adopted to
describe it.

Euclidean spacetime coordinates $x_0,x_1,x_2,x_3$, with the
metric~\mbox{$(g_{\mu\nu}) = {\rm diag}(1,1,1,1)$} are used.  The (vacuum)
fluctuating field is assumed to be an Abelian gauge field $A_\mu$, with
\mbox{$\mu = 0, 1, 2, 3$}, interacting with a zero-width mirror. This
interaction is realized, in a linear response approximation, via the VPT
due to a medium confined to a surface.  An approximation will be
implemented here, regarding this object: the surface curvatures shall be
assumed to be sufficiently small as to allow for a description where the
linear response function corresponding to a plane can be used locally. In
other words, at each tangent plane, we shall use the VPT due to a {\em
plane\/} mirror. The rationale behind this approximation, as well as the
kind of effect that are neglected by its implementation are discussed
below.

The most general kind of motion we shall deal with, amounts to a mirror
whose shape and position can be defined by a single scalar function $\psi$,
such that \mbox{$x_3=\psi(x_\parallel)$}, where $x_\parallel \equiv
(x_0,x_1,x_2)$.  Inside this general situation we shall focus on an
interesting particular case, namely, a situation where the mirror's surface
is defined by an equation of the form $x_3 = q(x_0)$,  
i.e., a rigidly moving infinite plane. The reason for considering this case
is that it will allow us to find explicit expressions for some interesting
models. Nevertheless, expressions corresponding to more general
functions will also be presented in the result for the effective action.

We would like to stress that, to make the calculations in Euclidean 
spacetime, is in no way mandatory; rather, it is a technique that, in some 
situations, may simplify the intermediate calculations without 
altering the underlying physics. For the particular problem
of moving mirrors,  the Euclidean formalism has already been used
in Ref.\cite{GK}. The connection with the ``in-out" and ``in-in" formalisms
has been discussed in detail in our previous work \cite{prd07}. For more
general discussions see Ref.\cite{vilkov}. 

The Euclidean action ${\mathcal S}$ is assumed to fall under the
general structure:
\begin{equation}\label{eq:defs}
	{\mathcal S} \;=\; {\mathcal S}(A,\psi) \,=\,  {\mathcal S}_0(A)
\,+\, {\mathcal S}_I(A,\psi) \;,
\end{equation}
where $S_0$ denotes the free action for the electromagnetic field
\begin{equation}\label{eq:defsa}
{\mathcal S}_0(A) \,=\, \frac{1}{4} \int F_{\mu\nu}F_{\mu\nu} \;,
\end{equation}
where $\mu, \nu$ and, in general, indices from the middle of the Greek
alphabet are assumed to run from $0$ to $3$.  On the other hand, ${\mathcal
S}_I(A,\psi)$ accounts for the coupling between $A$ and the microscopic
degrees of freedom on the moving mirror. To construct it, we start by
defining it in the case of a flat and static mirror at \mbox{$x_3 = \psi_0
= {\rm constant}$}:
\begin{equation}\label{eq:defsiflat}
	{\mathcal S}_I(A, \psi_0) \;=\; \frac{1}{2}
	\int_{x_\parallel,y_\parallel} A_\alpha(x_\parallel,\psi_0)
	\Pi^{\alpha\beta}(x_\parallel-y_\parallel)
	A_\beta(y_\parallel,\psi_0) \;,
\end{equation}
where $\Pi^{\alpha\beta}(x_\parallel-y_\parallel)$ denotes the VPT for the
medium on the plane sheet, and $\alpha, \beta = 0,1,2$. It depends, in
this case, on the difference between its arguments, because of the
(assumed) homogeneity of the medium.  We are working in the
usual linear response  approximation, in which one retains 
only the quadratic terms in the gauge field.

Note that the component of the gauge field which is normal to the mirror's
plane ($A_3$) does not couple to the medium, something which is perfectly
consistent with the assumption about the mirror to have zero width,
since in this case the current associated to the microscopic degrees of freedom is 
confined to the plane.

We consider now the general form of the interaction term
for a moving and deformed mirror described by $x_3=\psi(x_\parallel)$.
Formally, the integration of the microscopic degrees of freedom
must be performed in a curved hypersurface, whose
induced metric reads
\begin{equation}
	g_{\alpha\beta}(x_\parallel) \;=\; \delta_{\alpha\beta} +
	\partial_\alpha \psi(x_\parallel) \, \partial_\beta
	\psi(x_\parallel) \;.
\end{equation}
Therefore, on general grounds we expect the interaction term to 
have the covariant expression
\begin{eqnarray}\label{eq:defspsi}
{\mathcal S}_I(A,\psi) &=& \frac{1}{2} \,\int_{x_\parallel,y_\parallel}
\sqrt{g(x_\parallel)} \sqrt{g(y_\parallel)} \,  
A_\alpha(x_\parallel,\psi(x_\parallel))\nonumber \\
&\times &  \, 
\Pi_g^{\alpha\beta}(x_\parallel,y_\parallel) 
A_\beta(y_\parallel,\psi(y_\parallel))
\end{eqnarray}
where 
 $g(x_\parallel) = \det [g_{\alpha\beta}(x_\parallel)] = 1 +
\partial_\alpha \psi(x_\parallel) \partial_\alpha \psi(x_\parallel)$.

The components of the gauge fields which are parallel the mirror ($A_{\alpha}$) can be written
in terms of the components  in the laboratory frame ($A_\mu$) using
three tangent
vectors to the world-volume swept by the mirror during its time evolution:
\begin{equation}\label{eq:defemualpha}
	e^\mu_\alpha (x_\parallel) \,=\, \delta^\mu_\alpha \,+\,
	\delta^\mu_3\, \partial_\alpha \psi(x_\parallel)\;
\end{equation}
as follows
\begin{equation}
A_\alpha(x_\parallel)=e^\mu_\alpha (x_\parallel)A_\mu(x_\parallel).
\end{equation}
Note that, unlike for a flat and  static mirror,  the interaction term 
contains the laboratory frame component $A_3$.

$\Pi^{\alpha\beta}_g(x_\parallel,y_\parallel)$ in Eq. 
(\ref{eq:defspsi})
denotes the VPT for the medium
on the curved mirror (the subindex $g$  emphasizes the dependence of the VPT 
with the induced  metric). For an arbitrary $\psi$,  this is a rather 
involved object, which can be computed, in principle, using 
techniques of quantum field theory in curved spacetimes.
We shall assume that the induced metric 
is almost flat, i.e. $g_{\alpha\beta} \simeq \delta_{\alpha\beta}$. Physically,
this means that at each time  the mirror is gently curved, and that its motion
involves non-relativistic velocities, so that $\partial_\alpha \psi(x_\parallel) \, \partial_\beta
\psi(x_\parallel)\ll 1$.

We proceed as usual and expand the VPT around the flat metric
\begin{eqnarray}
\Pi^{\alpha\beta}_g(x_\parallel,y_\parallel)&\simeq& \Pi^{\alpha\beta}(x_\parallel-y_\parallel)
+ \Delta\Pi^{\alpha\beta}_g(x_\parallel,y_\parallel)\nonumber\\
&=&  \Pi^{\alpha\beta}(x_\parallel-y_\parallel)
+ O(\psi^2)
\, ,
\label{pi approx}
\end{eqnarray}
where the first term is the flat VPT and the first correction, linear in the metric, 
will be quadratic in $\psi$. 
One could compute  $\Delta\Pi^{\alpha\beta}_g(x_\parallel,y_\parallel)$, using, for instance
a covariant perturbation theory \cite{vilkov}. However,
as we will see, the flat VPT will be enough for our purposes.

Our next step is to introduce $\Gamma(\psi)$, the effective action for the mirror's
configuration:
\begin{equation}\label{eq:defgammapsi}
	e^{-\Gamma(\psi)} \,=\, \frac{{\mathcal Z}(\psi)}{{\mathcal Z}(0)} \;,
\end{equation}
where 
\begin{equation}\label{eq:defza}
	{\mathcal Z}(\psi)\,=\, \int [{\mathcal D}A] \; e^{- {\mathcal S}(A,\psi)}
	\;,
\end{equation}
and $[{\mathcal D}A]$ is the path integral measure including gauge fixing.

\subsection{Auxiliary fields}

Then we use an equivalent way of writing the $S_I$ term, by means of
an auxiliary field $\xi^\alpha(x_\parallel)$, a vector field in $2+1$
dimensions:
\begin{eqnarray}\label{eq:lagrange}
	e^{ -{\mathcal S}_I(A,\psi)} &=& {\cal N} \,\, \int {\mathcal D}\xi \, 
\delta(\nabla \cdot \xi)  \\ &\times & e^{ - \frac{1}{2} \int_{x_\parallel,y_\parallel}
\xi^\alpha(x_\parallel) \Lambda_{\alpha\beta}(x_\parallel,
y_\parallel) \xi^\beta(y_\parallel) + i \int_x J_\mu(x) A_\mu(x)} \nonumber 
\end{eqnarray}
where we introduced $J^\mu(x)$, a current concentrated on the mirror's
world-volume:
\begin{equation}\label{eq:defjmu}
	J^\mu(x) \,=\,  \sqrt{g(x_\parallel)} \, e^\mu_\alpha (x_\parallel)
	\, \delta(x_3-\psi(x_\parallel)) \, \xi^\alpha(x_\parallel) \;,
\end{equation}
and $\Lambda_{\alpha\beta}(x_\parallel, y_\parallel)$ is the inverse
(with respect to continuous and discrete indices) of $\Pi_g^{\alpha\beta}(x_\parallel, y_\parallel)$.
The inverse is understood on the space of fields satisfying
$\nabla_\alpha \xi^\alpha = 0$,  
the covariant
divergence of the auxiliary field. 
The reason for introducing the factor with a functional $\delta$-function
of this divergence has to do with the transverse nature of the vacuum
polarization tensor (Ward-Takahashi identities): the Gaussian
representation used in Eq.(\ref{eq:lagrange}) has to be constructed not with
unconstrained vector fields but rather only with transverse ones. Indeed,
$\Pi_g^{\alpha\beta}$ is invertible, and has $\Lambda_{\alpha\beta}$ as its
inverse, on the subspace of transverse fields.  Besides, the condition on
the divergence of the auxiliary field implies the conservation of $J^\mu$,
and hence the invariance of the action under gauge transformations $A_\mu
\to A_\mu + \partial_\mu \omega$. 

Using now Eq.(\ref{eq:lagrange}) in Eq.(\ref{eq:defza}), and integrating out $A$, we see that:
\begin{equation}
	{\mathcal Z}(\psi) \,=\, {\mathcal Z}_0 \,
	{\cal N}\, 
	\int {\mathcal D}\xi \, \delta(\nabla \cdot \xi) \, 
	e^{ -{\mathcal S}_{\rm eff}(\xi)} 
\end{equation}
where ${\mathcal Z}_0$ is the vacuum amplitude for the free electromagnetic
field, and
\begin{equation}\label{eq:defseff}
	{\mathcal S}_{\rm eff}(\xi) \,=\, \frac{1}{2}
	\int_{x_\parallel,y_\parallel}
	\xi^\alpha(x_\parallel) {\mathcal K}_{\alpha\beta}(x_\parallel,y_\parallel) 
	\xi^\beta(y_\parallel)\;,
\end{equation}
where
\begin{eqnarray}
	{\mathcal K}_{\alpha\beta}(x_\parallel,y_\parallel) &=&
	\sqrt{g(x_\parallel)} \,e^\mu_\alpha (x_\parallel) \,
	G_{\mu\nu}\big(x_\parallel-y_\parallel,\psi(x_\parallel)-\psi(y_\parallel)\big) \nonumber \\
	&\times &e^\nu_{\beta}(y_\parallel) \,\sqrt{g(y_\parallel)} 
+ \Lambda_{\alpha\beta}(x_\parallel, y_\parallel) \;,
\label{mathcalK}
\end{eqnarray}
with $G_{\mu\nu}(x-y)$ denoting the gauge field propagator. 
We have found it convenient to use the Feynman gauge, so that
\begin{equation}
	G_{\mu\nu}(x-y) \;=\;\delta_{\mu\nu} \int \frac{d^4k}{(2\pi)^4} \, 
\frac{e^{i k\cdot (x-y)}}{k^2} \;.
\end{equation}	
Note that, since the auxiliary field is constrained to verify $\nabla \cdot
\xi =0$,  in Eq.(\ref{eq:defseff}) we may discard from ${\mathcal
K}_{\alpha\beta}$ any contribution which vanishes when acting on the
subspace of fields satisfying that condition.

\subsection{Second order expansion}

Let us now implement the second order perturbative expansion for the
effective action $\Gamma(\psi)$ to the second order in $\psi$, the
departure of the mirror from its average position.
We first note that the formal result of integrating out the auxiliary field is:
\begin{equation}
	\Gamma(\psi) \,=\, \frac{1}{2} \,\Big[ 
		{\rm Tr} \log {\mathcal K} \,-\, {\rm Tr} \log {\mathcal
	K}|_{\psi \equiv 0}\Big] \;.
\end{equation}
Denoting by ${\mathcal K}^{(a)}$ the $a^{th}$-order term in an expansion in
powers of $\psi$, we obtain the corresponding expansion of $\Gamma$. The
$0^{th}$-order term vanishes, as well as the first-order term, while the
second-order term $\Gamma^{(2)}$ becomes:
\begin{equation}
	\Gamma^{(2)}(\psi) \,=\, \frac{1}{2} \,{\rm Tr} \Big[ \big({\mathcal
	K}^{(0)}\big)^{-1} {\mathcal K}^{(2)}\big] \;.
\end{equation}

In this approximation, the effective action will be of the form
\begin{equation}
	 \Gamma^{(2)} (\psi)\;=\; -\frac{1}{2} \int _{x_\parallel,x'_\parallel} \, \psi(x_\parallel)
	F(x_\parallel,x'_\parallel)   \psi(x'_\parallel)
\end{equation}
for some two-point function $F$. As we are interested  in dissipative effects,
we may neglect any contribution to $ \Gamma^{(2)}$ which is local in derivatives
of $\psi$.
This will simplify the calculations below.

The explicit form of the zeroth-order kernel, which is translation invariant, is 
\begin{equation}
	{\mathcal K}^{(0)}_{\alpha\beta}(x_\parallel,y_\parallel) \,=\,
	\int \frac{d^3k_\parallel}{(2\pi)^3} \, e^{i k_\parallel \cdot
	(x_\parallel-y_\parallel)} \, 	\widetilde{\mathcal
	K}^{(0)}_{\alpha\beta}(k_\parallel) 
\end{equation}
where
\begin{equation}
\widetilde{\mathcal K}^{(0)}_{\alpha\beta}(k_\parallel) \,=\,
	\frac{1}{2 |k_\parallel|} 
	{\mathcal P}^\perp_{\alpha\beta}(k_\parallel) \,+\,
	\widetilde{\Lambda}^{(0)}_{\alpha\beta}(k_\parallel) \;,
\end{equation}
$\widetilde{\Lambda}^{(0)}_{\alpha\beta}$ is the Fourier transform
of the inverse of the flat VPT and 
${\mathcal P}^\perp_{\alpha\beta} \equiv \delta_{\alpha\beta} -
\frac{k_\alpha k_\beta}{k_\parallel^2}$.

Regarding the second-order object ${\mathcal K}^{(2)}$, it receives many
different contributions. However, as already stressed, we are here interested only in the
calculation of dissipative terms. Thus, we may ignore any term producing
a local contributions.
In particular, as the deviation of the induced metric from the identity tensor is 
already quadratic in $\psi$,  we can replace it by the identity tensor.
For the same reason, we 
can omit the corrections to the flat VPT in Eq.(\ref{pi approx}), and
replace $\Lambda_{\alpha\beta}$ by $\Lambda^{(0)}_{\alpha\beta}$
in Eq.(\ref{mathcalK}).
Note however, that there will be a non trivial contribution from the tangent vectors $e_\alpha^\mu$,
that contain terms linear in $\psi$. 
Thus,
\begin{eqnarray}
{\mathcal K}^{(2)}_{\alpha\beta}(x_\parallel,y_\parallel) &=&
-\frac{1}{2} \psi(x_\parallel) \psi(y_\parallel) \,
\int \frac{d^3k_\parallel}{(2\pi)^3} \, e^{i k_\parallel \cdot (x_\parallel-y_\parallel)} \, 
|k_\parallel| \delta_{\alpha\beta} \nonumber\\
&+& \partial_\alpha\psi(x_\parallel) \partial_\beta\psi(y_\parallel) \,
\int \frac{d^3k_\parallel}{(2\pi)^3} \, \frac{e^{i k_\parallel \cdot
(x_\parallel-y_\parallel)}}{2 |k_\parallel|} \;.
\end{eqnarray}

We may obtain a more explicit formula for the second order contribution to
$\Gamma$, by taking into account the structure of the VPT, which appears in
$\widetilde{\mathcal K}^{(0)}$. Under the assumption of invariance under
spatial rotations on $x_3$=constant planes, this tensor can be decomposed
into orthogonal projectors. 
Indeed, since:
\begin{equation}\label{eq:trans}
	k_\alpha \widetilde{\Pi}^{\alpha \beta}\;=\; 0 \;,
\end{equation}
the irreducible tensors (projectors) along which
$\widetilde{\Pi}_{\alpha\beta}$ may be decomposed must satisfy the
condition above and may be constructed using as building blocks the objects: $\delta_{\alpha\beta}$,
$k_\alpha$, and \mbox{$n_\alpha = (1,0,0)$}.  By performing simple
combinations among them, we also introduce: \mbox{$\breve{k}_\alpha \equiv
k_\alpha - k_0 n_\alpha$}, and  \mbox{$\breve{\delta}_{\alpha\beta} \equiv
	\delta_{\alpha\beta} - n_\alpha n_\beta$}. 

Then we construct two independent tensors satisfying the transversality
condition, ${\mathcal P}^t$ and ${\mathcal P}^l$, defined as follows:
\begin{equation}
{\mathcal P}^t_{\alpha\beta} \,\equiv\, \breve{\delta}_{\alpha\beta} -
\frac{\breve{k}_\alpha\breve{k}_\beta}{\breve{k}^2}
\end{equation}
and 
\begin{equation}
{\mathcal P}^l_{\alpha\beta} \,\equiv\, {\mathcal
P}^\perp_{\alpha\beta}\,-\, {\mathcal P}^t_{\alpha\beta} \;.
\end{equation}
Defining also:
\begin{equation}
{\mathcal P}^\shortparallel_{\alpha\beta} \,\equiv\, \frac{k_\alpha k_\beta}{k^2} \;.
\end{equation}
we find the following algebraic properties:
	$$
	{\mathcal P}^\perp + {\mathcal P}^\shortparallel = I \;,\;\;
	{\mathcal P}^t + {\mathcal P}^l = {\mathcal P}^\perp 
	$$
	$$
	{\mathcal P}^t {\mathcal P}^l = {\mathcal P}^l {\mathcal P}^t = 0
	\;,\;\; {\mathcal P}^\shortparallel {\mathcal P}^t = {\mathcal P}^t {\mathcal P}^\shortparallel = 0 
	\;,$$
	$$\;\; {\mathcal P}^\shortparallel {\mathcal P}^l = {\mathcal P}^l {\mathcal P}^\shortparallel = 0 \;,
	$$
\begin{equation}
\big({\mathcal P}^\perp\big)^2 = {\mathcal P}^\perp \;,
\big({\mathcal P}^\shortparallel\big)^2 = {\mathcal P}^\shortparallel \;,
\big({\mathcal P}^t\big)^2 = {\mathcal P}^t \;,
\big({\mathcal P}^l\big)^2 = {\mathcal P}^l \;.
\end{equation}

For a  general medium, we shall have:
\begin{equation}
	\widetilde{\Pi}_{\alpha\beta}(k) \,=\,  
	g_t\big(k_0, {\mathbf k}_\parallel\big) \, {\mathcal P}^t_{\alpha\beta} \,+\, 
	g_l\big(k_0, {\mathbf k}_\parallel\big) \, {\mathcal P}^l_{\alpha\beta}
	\;,
	\label{piscalar}
\end{equation}
where $g_t$ and $g_l$ are model-dependent scalar functions.

In what follows, we particularize to the case of the rigid motion of a
flat mirror along its normal direction, i.e. $\psi(x_\parallel)=q(x_0)$. 
In this case $\Gamma^{(2)}$ has the form:
\begin{eqnarray}
	\frac{1}{L^2} \Gamma^{(2)} \;&=&\; -\frac{1}{2} \int _{x_0,x'_0} \, q(x_0)
	f(x_0-x'_0)   q(x'_0)\nonumber\\
	&=& \frac{1}{2}\int dp_0 {\tilde
f}(p_0)\vert\tilde q(p_0)\vert^2\, ,
\label{gamma2fin}
\end{eqnarray}
where $L^2$ denotes the area of the $x_1,x_2$ space. Using the projectors
introduce above, after some algebra, we see that ${\tilde f}(p_0)$, the
Fourier transform of $f$, naturally decomposes as follows: 
\begin{equation}\label{eq:fdecomp}
{\tilde f}(p_0) \,=\, {\tilde f}_t(p_0) + {\tilde
f}_l(p_0) 
\end{equation}
where
\begin{equation}\label{eq:fnu2}
	{\tilde f}_t(p_0) \;=\; \int \frac{d^3k_\parallel}{(2\pi)^3} \; r_t(k_\parallel) \, 
	|k_\parallel| \, \sqrt{(k_0 + p_0)^2 + {\mathbf k_\parallel}^2} \;, 
\end{equation}
and
\begin{eqnarray}\label{eq:fnu1}
{\tilde f}_l(p_0) &=&  \int \frac{d^3k_\parallel}{(2\pi)^3} \;
r_l(k_\parallel) \,\left[ |k_\parallel| \, 
\sqrt{(k_0 + p_0)^2 + {\mathbf k_\parallel}^2}\right. \nonumber\\
&-& \left.\frac{{\mathbf k_\parallel}^2 \, p_0^2}{|k_\parallel|\sqrt{(k_0 + p_0)^2 + {\mathbf k_\parallel}^2}  }
 \right]
\;,
\end{eqnarray}	
where
\begin{equation}
r_{t,l}(|k_\parallel|) \,=\, 
		\frac{1}{ 1 + \frac{2
|k_\parallel|}{g_{t,l}(k_\parallel)}} \;.
\label{reflcoef}
\end{equation}

The decomposition in Eqs.(\ref{eq:fdecomp})-(\ref{eq:fnu1}), the main general
result of this article, implies that the EM field problem is decomposed
into two independent contributions, one due to $r_t$ and the other to
$r_l$, which, as we will see below,  are the Euclidean version of the TE and 
TM mirror's reflection coefficients.

The dissipative effects can be obtained from the imaginary part of the real
time ``in-out" effective action $\Gamma^{(2)}_{\rm in-out}$, which is
related to the probability $P$ of producing
a photon pair out of the vacuum \cite{vpa} through
\begin{equation}
P\simeq 2\, {\rm Im}[\Gamma^{(2)}_{\rm in-out}] \, .
\end{equation}
The ``in-out" effective action can be obtained from the Euclidean effective action performing a
Wick rotation. From Eq.(\ref{gamma2fin}) we obtain
\begin{equation}
\frac{\Gamma^{(2)}_{\rm in-out}}{L^2} =\frac{1}{2}\int dp_0 [{\tilde
f}_{t}(ip_0)+{\tilde f}_l(ip_0)]\vert\tilde q(p_0)\vert^2\, ,
\end{equation}
where  $\tilde q(p_0)$ denotes the Fourier transform of the physical trajectory
in Minkowski spacetime.

It is noteworthy that the coefficients $r_{t,l}$ appearing in Eq.(\ref{reflcoef}) and
in  the final
formula for the effective action are just the static TE and TM reflection coefficients of the
mirror. Indeed, in Ref.~\cite{fialkovsky} it has been shown that for a thin
mirror characterized by its VPT,  the Euclidean reflection coefficients are given
by
\begin{equation}
r_{TM}=\frac{1}{1+\frac {2{\mathbf k}^2_\parallel}{\vert k_\parallel\vert\widetilde\Pi_{00}}}\label{rtertm0}\end{equation}
\begin{equation} r_{TE} = \frac{-k_\parallel^2 \, \widetilde\Pi_{00} + {\mathbf k}^2_\parallel\, \widetilde \Pi_{\alpha\alpha} }{-k_\parallel^2\,  \widetilde\Pi_{00} + {\mathbf k}^2_\parallel\, 
 \widetilde \Pi_{\alpha\alpha}+
 2\vert k_\parallel\vert{\mathbf k}^2_\parallel}\, .
 \label{rtertm}
 \end{equation}
 From Eq.(\ref{piscalar}) and the definitions of the transverse and longitudinal projectors it is easy to see that
 \begin{eqnarray}
 \widetilde\Pi_{00}&=& \frac{{\mathbf k}^2_\parallel}{k_\parallel^2}g_l\nonumber\\
 \widetilde \Pi_{\alpha\alpha}&=&g_l+g_t. 
 \end{eqnarray}
 Inserting these results into Eqs.(\ref{rtertm0}) and (\ref{rtertm}) one can verify that $r_t=r_{TE}$ and $r_l=r_{TM}$.

An important remark is in order here, namely, that there are still in
$\tilde f$, as given by (\ref{eq:fnu2}), contributions that are cancelled
by the subtraction of $\psi=0$, time-independent effects. In Fourier space
we shall simply implement that by subtracting from ${\tilde f}$ its value
at $p_0 =0$. Besides, depending on the large momentum behaviour of the
$r_{t,l}$ functions, we may also need to perform the subtraction of more
terms inside the momentum integral. Indeed, depending on the superficial
degree of divergence, we shall need to subtract from the integrand a
polynomial of a higher degree in $p_\parallel$. Note that these terms will
not affect the imaginary part of the effective action, and could be
absorbed by redefining the mass and, eventually, terms with higher
derivatives in the classical action for the mirror.

In what follows we evaluate the resulting $\tilde{f}$ and its analytic continuation for some
interesting examples.
\section{Examples}\label{sec:examples}

\subsection{The thin  perfect conductor}

We start by considering the simplest case of perfect conductivity, for which $r_{t,l}=1$.

To evaluate the contribution of the TE mode, we need to evaluate the integral
\begin{equation}
	{\tilde f}_t(p_0) \;=\;  \int \frac{d^3k_\parallel}{(2\pi)^3} \;
	|k_\parallel| \, \sqrt{(k_0 + p_0)^2 + {\mathbf
k_\parallel}^2}\;,
\end{equation}
or, using spherical coordinates
\begin{equation}
	{\tilde f}_t(p_0) = \frac{1}{4\pi^2}\int_0^\infty dk\int_0^\pi d\theta k^3 \sin\theta
	\sqrt{k^2+p_0^2+2 k p_0\cos\theta}.
\end{equation}

This integral is of course divergent. As mentioned in the previous section, to
renormalize the form factor $\tilde
f_t$ we follow a BPHZ approach, subtracting from the integrand its
Taylor expansion in $p_0^2$ around $0$, up to the order $p_0^4$ (dictated
by the superficial degree of divergence).

After the subtraction, the integrals in $\theta$ and $k$ can be performed analytically \cite{math}. The result is
\begin{equation}
	{\tilde f}_t(p_0) \;=\; -\frac{\vert p_0\vert^5}{360\pi^2}
	\label{TEperf}
\end{equation}
which is the well known result for TE contribution \cite{paulo94}.

We now consider the TM contribution. Using again spherical coordinates, the form factor is 
given by
\begin{eqnarray}
	{\tilde f}_l(p_0) &=& \frac{1}{4\pi^2}\int_0^\infty dk\int_0^\pi d\theta k^3 \sin\theta
	\sqrt{k^2+p_0^2+2 k p_0\cos\theta}\nonumber\\
	&\times& \left(1-\frac{p_0^2 \sin^2\theta}{k^2+p_0^2+2 k p_0\cos\theta}\right) .
\end{eqnarray}
Following the same steps as before we obtain
\begin{equation}
	{\tilde f}_l(p_0) \;=\; -\frac{11\vert p_0\vert^5}{360\pi^2}\,
\end{equation}
which reproduces the TM contribution computed using different methods \cite{paulo94}.

The perfect conductor limit is useful not only as a consistency check of our calculations. Indeed, as pointed out in Ref. \cite{Bordag},
there is a subtle difference in the boundary conditions for thin and thick perfect conductors. Although this difference is not manifested in 
the static Casimir effect, it influences the Casimir-Polder interaction. We have seen that this is not the case for the DCE.

\subsection{A medium with constant functions $g_{t,l}$}

We shall consider here the case in which the functions
$g_{t,l}$ are constant; namely, $g_{t,l} = \lambda_{t,l}$. The calculation can be performed
following the same steps than for the perfect conductor case, inserting
into the integrals the reflection coefficients
\begin{equation}
r_{t,l}(|k_\parallel|) \,=\, 
		\frac{1}{ 1 + \frac{2
|k_\parallel|}{\lambda_{t,l}}} \;.
\end{equation}

It is worth to stress that the expression for the TE
contribution coincides exactly with that of a
quantum scalar field with a $\delta$-potential, that was considered in
Ref.\cite{prd07}. Therefore, a VPT with a constant functions $g_t$ yields,
for the TE mode, the natural EM generalization of the scalar problem, that was considered in
several previous works to analyze the static and DCE \cite{milton}.

Using again spherical coordinates, we first subtract the Taylor expansion up to the 
order $p_0^4$, and then compute the integral in $\theta$.
The resulting expression can be
integrated in $k$ analytically. The result is:
\begin{equation}
{\tilde f}_t(p_0) \;=\;- |p_0|^5 \, \varphi_t\Big(\frac{|p_0|}{\lambda_t}\Big)
\label{ft},
\end{equation}
where
\begin{eqnarray}
\varphi_t(\xi) &=& \frac{24 \xi ^5-60 \xi ^4-220 \xi ^3-150 \xi ^2-30 \xi} {5760 \pi ^2 \xi ^6} \nonumber \\ 
&+& \frac{15
(2 \xi +1)^3 \log (2 \xi +1)}{5760 \pi ^2 \xi ^6} \;.
\end{eqnarray}

In the strong coupling (almost perfectly conducting mirror) limit, we
get the expansion:
\begin{equation}
{\tilde f_t}(p_0) \;=\; -\frac{|p_0|^5}{360 \pi ^2}+\frac{|p_0|^6}{420 \pi
^2\lambda_t} - \frac{|p_0|^7}{420 \pi ^2\lambda_t^2}
 + \ldots
\end{equation}
Among these terms, only the ones involving odd powers of $|p_0|$
contribute, when continued to real time, to the imaginary part of the
effective action:
\begin{equation}
{\rm Im} \Big[{\tilde f_t}(i p_0)\Big] \;=\; \frac{|p_0|^5}{360 \pi ^2} -
\frac{|p_0|^7}{420 \pi ^2\lambda_t^2} + \ldots
\end{equation}
where we recognize the leading term as identical to the one for a scalar field with Dirichlet
boundary conditions, and to the one of the TE modes of the electromagnetic field for perfect conductors.

In the weak coupling limit, we obtain the expansion:
\begin{eqnarray}
&& {\tilde f_t}(p_0) = 
-\frac{|p_0|^4 \lambda_t}{240 \pi ^2}+\frac{|p_0|^3 \lambda_t^2}{96 \pi ^2}\nonumber \\ &-&
\frac{|p_0|^2\left(-11 +6 \log(2 |p_0|)-6 \log\lambda_t\right) \lambda_t^3}{288 \pi ^2}
+ \ldots
\end{eqnarray}
where we neglected terms of higher order in $\lambda_t$. Therefore we obtain,  to leading order
\begin{equation}
{\rm Im} \Big[{\tilde f_t}(i p_0)\Big] \;=\; \frac{|p_0|^3 \lambda_t^2}{96 \pi ^2}\, .
\end{equation}

The form factor associated to the TM reflection coefficient can be computed along the same lines. The result can be written as
\begin{equation}
{\tilde f}_l(p_0) \;=\;- |p_0|^5 \, \varphi_l\Big(\frac{|p_0|}{\lambda_l}\Big)
\label{fl},
\end{equation}
where
\begin{eqnarray}
\varphi_l(\xi) &=& \frac{-872 \xi ^5+1020 \xi ^4+ 20 \xi ^3-690 \xi ^2-210 \xi} {28800 \pi ^2 \xi ^6} \nonumber \\ 
&+& \frac{15
(2 \xi +1)^3 (8\xi^2-12\xi+7) \log (2 \xi +1)}{28800 \pi ^2 \xi ^6} \;.
\end{eqnarray}
In the strong coupling limit we obtain
\begin{equation}
{\tilde f_l}(p_0) = -\frac{11 \vert p_0\vert^5}{360 \pi ^2}+\frac{\vert p_0\vert ^6}{60 \pi ^2 \lambda_l}-\frac{17 \vert p_0\vert^7}{1260 \pi ^2 \lambda_l^2}+
\dots
\end{equation}
that reproduces the perfect conductor result for $\lambda_l\to\infty$.

\begin{figure}[h!]
\includegraphics[width=\linewidth]{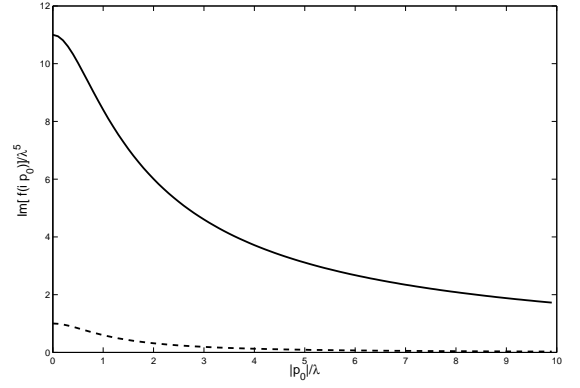}
\caption{${\rm Im}[{\tilde f_t(i p_0)}]/{\lambda}^5$ divided by the TE-perfect conductor result, as a function of  $|p_0|/\lambda$ for a rigid motion in a medium with constant $g_t=\lambda$ in the dashed line. The solid line plot is the ${\rm Im}[{\tilde f_l(i p_0)}]/{\lambda}^5$ divided by the TE-perfect conductor result, as a function of  $|p_0|/\lambda$. While dashed line coincides with the TE perfect conductor result for small $\vert p_0\vert/\lambda$, the solid line goes to 11, which is the correct limit \cite{paulo94}.}
\label{fig1}
\end{figure}

In Fig. \ref{fig1}  we plot the imaginary part ${\rm Im} \Big[{\tilde f}(i p_0)\Big]$ divided by  
the TE-perfect conductor result, as a function of the external frequency.  Solid line represents the result 
of Eq.(\ref{ft}) that, in the limit of $\vert p_0\vert/\lambda \rightarrow 0$, goes to 11, as it is remarked in Ref.\cite{paulo94}. Dashed line corresponds to  
Eq.(\ref{fl})  and it approaches 
to 1 in the zero frequency limit (coincides with the perfect conductor limit for the TE mode). Dissipative effects
grow with $p_0$. 

\subsection{Evaluation of $\tilde{f}(p_\parallel)$ for graphene}

For the case of graphene,  we may apply the tools introduced in the previous
section to decompose the vacuum polarization tensor in terms of the
irreducible projectors: 
\begin{equation}
	\widetilde{\Pi}_{\alpha\beta}(k) =\frac{e^2 N |m|}{4\pi}  
	F\big(\frac{k_0^2 + v_F^2 {\mathbf k}^2}{4 m^2}\big) 
	\Big[ {\mathcal P}^t_{\alpha\beta} + 
	\frac{k_0^2+{\mathbf k}^2}{k_0^2+ v_F^2 {\mathbf k}^2} {\mathcal P}^l_{\alpha\beta} 
	\Big]
\end{equation}
where:
\begin{equation}
	F(x) \;=\;   1 - \frac{1 - x}{\sqrt{x}} \, \arcsin[ ( 1 +
	x^{-1})^{-\frac{1}{2}}]\;,  
\end{equation}
$m$ is the mass (gap), $N$ the number of $2$-component Dirac fermion
fields, and $v_F$ the Fermi velocity (in units where $c=1$).

Usually, the most relevant case corresponds to $m=0$; when that is the case:
\begin{eqnarray}
	\widetilde{\Pi}_{\alpha\beta} &=& \frac{e^2 N}{16} \, 
	\sqrt{k_0^2 + v_F^2 {\mathbf k}^2}\; 
	\Big[ {\mathcal P}^t_{\alpha\beta} \,+\, 
	\frac{k_0^2+{\mathbf k}^2}{k_0^2+ v_F^2 {\mathbf k}^2} {\mathcal P}^l_{\alpha\beta} 
	\Big] \nonumber\\
                         &=& \frac{e^2 N}{16} \, \sqrt{k_0^2+ {\mathbf
			 k}^2} \,  
	\Big[ \, \sqrt{\frac{k_0^2 + v_F^2 {\mathbf k}^2}{k_0^2 +
	{\mathbf k}^2}} \;
	{\mathcal P}^t_{\alpha\beta}\nonumber \\  &+& 
	\sqrt{\frac{k_0^2 + {\mathbf k}^2}{k_0^2 + v_F^2
	{\mathbf k}^2}} \; {\mathcal P}^l_{\alpha\beta} 
	\Big] \;.
\end{eqnarray}

Then, coming back to the general formulae, we see that (with massless fermions), the 
reflection coefficients are 
\begin{equation}\label{eq:Gtg}
r_t(k_\parallel) \, = \, 
\frac{1}{1 + \frac{32}{e^2 N}\sqrt{\frac{k_0^2+{\mathbf k_\parallel}^2}{k_0^2+ v_F^2 {\mathbf
k_\parallel}^2}}} 
\end{equation}
and
\begin{equation}\label{eq:Glg}
r_l(k_\parallel) \, = \, \frac{1}{1 + \frac{32}{e^2 N}\sqrt{\frac{k_0^2+ v_F^2{\mathbf
k_\parallel}^2}{k_0^2+ {\mathbf k_\parallel}^2}}} \;.
\end{equation}

As there are no dimensionful constants in the VPT, dimensional analysis implies that
\begin{equation}
{\tilde f}( p_0) \;=\;  |p_0|^5 C(N e^2,v_F)\, ,
\label{fgraph}
\end{equation}
that is, the result is proportional to that of a perfect conductor.  The dimensionless function $C$
depends on the coupling constant and the Fermi velocity.

\begin{figure}[h!]
\includegraphics[width=\linewidth]{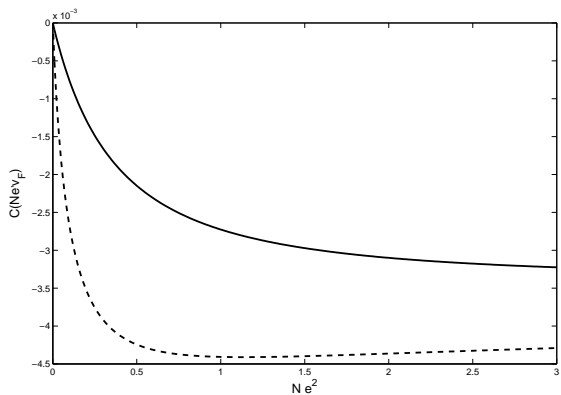}
\caption{$C(N e^2,v_F)$ as a function of the dimensionless coupling constant $N e^2$ for two different 
Fermi velocities $v_F = 0.5$ in the solid line plot, and $v_F= 0.05$ for the dashed line (in units in which $c=1$).
The form factors vanish  in the weak coupling limit $N e^2\to\ 0$, and 
 tend slowly to the perfect conductor limit $-1/(30 \pi^2)$
as $N e^2\to\infty$
}
\label{fig2}
\end{figure}

In order to compute explicitly this function, we insert the graphene reflection coefficients
into Eqs.(\ref{eq:fnu2}) and (\ref{eq:fnu1}), subtract the Taylor expansion up to order $p_0^4$,
and evaluate the integrals using spherical coordinates. Unlike the previous examples, 
the complicated dependence of the reflection coefficients with the angle $\theta$
makes not possible to compute analytically this integral. Therefore we computed 
the form factors numerically for different values of the coupling constants and Fermi velocity.
The results are shown in Fig. \ref{fig2}. As expected, the form factors tend to the perfect conductor limit $C\to-1/(30\pi^2)$
as $N e^2\to\infty$, and vanish  in the weak coupling limit $N e^2\to\ 0$. Note that, for small values of the Fermi velocity,
the behaviour is non-monotonous with the coupling constant. Most notably, for some values of the parameters,
the dissipative effects can be larger for graphene than for perfect conductors. 

In the particular case $v_F \to 1$, relativistic fermion limit, the reflection coefficients 
become constants, and the results for the TE and TM  form factors are those of the perfect conductor
divided by the factor $1+32/(e^2 N)$.

We end this section pointing out that
it would be interesting to compute the VPT in the framework
of quantum fields in curved spaces, beyond the weak field approximation, 
and check explicitly the validity of the approximation 
$\Pi^{\alpha\beta}_g\simeq \Pi^{\alpha\beta}$ used in Eq.(\ref{pi approx}).

\section{Conclusions}\label{sec:conclusions}

We have obtained a general expression for the effective action
corresponding to a single imperfect mirror coupled to the EM field, to
second order in the departure of the mirror from its equilibrium position.
The resulting formula decomposes into two scalar like contributions, in
terms of two scalar functions that define the VPT.  The final expression
for the effective action can be written in a very compact way in terms of
the TE and TM reflection coefficients of the mirror.  These results can be
considered as a generalization to the electromagnetic case of those in
Ref.\cite{prd07}, where we considered scalar and spinorial vacuum fields and modeled the
interaction between an imperfect mirror and the vacuum field using  a
$\delta$-potential.  

We have evaluated explicitly the effective action for some examples, which
in our context correspond to the use of the corresponding VPT. We have
obtained the vacuum decay amplitude using a proper analytic continuation of
the Euclidean results.  

We have shown that our results reproduce correctly
the TE and TM contributions in the case of perfect conductors.
For the particular case of graphene,  we have shown that the imaginary
part of the effective action is that of a perfect conductor times a function
that depends on the coupling constant and the Fermi velocity. We computed 
explicitly this function and found a non-monotonous behavior with the coupling constant.
Moreover, for some values of
the parameters, the dissipative effects may be larger than those for  a perfect conductor.

It would be of interest to compute the VPT beyond the weak field approximation,
in particular for the case
of massless fermions. We hope to address this 
relevant issue in a forthcoming work.  This kind of system would require
a fuller knowledge of the dependence of the VPT on the geometry. Even in the
absence of coupling to the gauge field, a curved monolayer graphene 
can be considered as a physical realization of quantum field theory
in curved spacetimes \cite{Iorio}, providing condensed matter analogues
of semiclassical gravitational effects. 



\begin{thebibliography}{99}
\bibitem{vpa} More generally,  $P$ is the total vacuum decay probability, which also includes the possibility of excitation of the internal degrees of freedom of the mirror. These internal degrees of freedom also produce inertial forces and dissipation on the 
accelerated mirror, see for instance Ref.\cite{inertial}.
\bibitem{math} We computed the integrals using {\it Mathematica}.  A useful point is to note that, after the integration in $\theta$, the resulting function vanishes 
  for $k>\vert p_0\vert$.
\end{thebibliography}
\end{document}